\documentclass[12pt]{article}
\usepackage{latexsym, epsfig, graphics, amsmath,}
\numberwithin{equation}{section}
\newcommand{\be}{\begin{equation}}
\newcommand{\ee}{\end{equation}}
\newcommand{\bq}{\begin{eqnarray}}
\newcommand{\eq}{\end{eqnarray}}
\newcommand{\cue}{{\bf q}}
\newcommand{\dbsm}{\sum_{\sigma_{1}<\sigma}\sum_{\sigma<\sigma_{2}}}
\newcommand{\sqsm}{\sqrt{(\sigma-\sigma_{1})(\sigma_{2}-\sigma_{1})
(\sigma_{2}-\sigma)}}
\newcommand{\wss}{W(\sigma_{1},\sigma_{2})}
\newcommand{\ssm}{\frac{\sigma_{2}-\sigma_{1}}{(\sigma_{2}-\sigma)
(\sigma -\sigma_{1})}}
\newcommand{\bfv}{{\bf v}}
\newcommand{\qsm}{(\sigma -\sigma_{1}) (\sigma_{2} -\sigma_{1})
(\sigma_{2} -\sigma)}
\begin{document}
\begin{titlepage}
\today          \hfill 
\begin{center}

\vskip .5in

{\large \bf Scalar Field Theories On The World Sheet: A Non-Trivial Ground
State }\footnote{This work was supported by the Director, Office of
Science, Office of High Energy Physics of the U.S. Department of Energy
under Contract DE-AC02-05CH11231.}

\vskip .50in


\vskip .5in
Korkut Bardakci \footnote{Email: kbardakci@lbl.gov}

{\em Department of Physics\\
University of California at Berkeley\\
   and\\
 Theoretical Physics Group\\
    Lawrence Berkeley National Laboratory\\
      University of California\\
    Berkeley, California 94720}
\end{center}

\vskip .5in

\begin{abstract}

The present article completes an earlier publication, which was the
 culmination of a series of papers dedicated to
 the study of the planar graphs
of the scalar $\phi^{3}$ theory on a light cone world sheet. 
 In the earlier work, a  field theory on
a continuous world sheet that reproduces these planar graphs
 was constructed, and the mean field approximation
was applied to it. This led to the formation of a soliton, and the
fluctuations around the soliton were identified with 
 stringy excitations.
We point out, however, that in this earlier work, a complete treatment of the
ground state of the model was missing. This was due to an unnecessary
decompactification of the world sheet; by keeping it compactified,
we show that, in addition to a trivial ground state, there is also
a non-trivial one. We investigate fluctuations around the non-trivial
ground state in the limit of a densely populated world sheet, and show
string formation in this limit. We also show that this limit can be
systematically studied by means of an expansion in terms of a
conveniently defined coupling constant.

\end{abstract}
\end{titlepage}

\newpage
\renewcommand{\thepage}{\arabic{page}}
\setcounter{page}{1}
\section{Introduction}
\vskip 9pt

The present work can be thought of as a supplement to a previous
article [1]: It completes that article by providing a crucial final
step that was missing. We could have written a short note on
just this final step, but instead we decided on a longer article
  that aims to be
self contained for the  convenience of the reader. Sections 1 through
6 are essentialy a rewrite of [1] with only a few minor modifications
and a  section deleted.
The reader who is only interested in the new material could skip directly
to section 7.

 Reference [1] was the culmination of a long
development starting with [2]. The idea behind this program was to sum the 
planar graphs of a
 field theory on a world sheet parametrized by the
light cone variables, based on 't Hooft's pioneering paper [3]. The
original field theory studied in this approach was a scalar with
$\phi^{3}$ interaction, and this was later generalized to more
complicated and more interesting models [4, 5]. The model under
consideration here is again scalar $\phi^{3}$ in transverse
dimensions $D=1, 2, 4$. For the sake of brevity,  section 7 in [1],
where an additional $\phi^{4}$ interaction was introduced, has been
omitted.

The starting point  is the world sheet field theory, 
 which reproduces the planar graphs of $\phi^{3}$ [1]. 
This theory is based on a complex scalar field and a two component
fermion field that live on the world sheet.  Using the mean field
approximation, solitonic classical solutions
on the world sheet were
constructed, and a certain set of quantum fluctuations about the
solitonic solutions were shown to have a string like spectrum.
The solitonic solutions are of interest because they describe a 
non-perturbative feature of field theory. Also, as we shall see later,
the soliton emerges from  the summation of a dense set of graphs
on the world sheet, which can be thought of as the condensation of
these graphs. The existence of such a condensate on the world sheet
is naturally expected to lead to a string description, an old idea
that motivated some of the early work on this subject [6, 7].

These computations  suffer from two kinds of
divergences: One of them is the  field theoretic ultraviolet
divergences, which are eliminated by the standard renormalization
procedure.
  The second one is a spurious 
 infrared divegence due to the choice of the light cone
coordinates. In the previous work, this infrared problem was temporarily
 avoided by the discretizing the $\sigma$ coordinate on the world sheet
in steps of $a$, but then, several quantities of physical interest were
 singular  in the limit
$a\rightarrow 0$. The main reult of [1] is that this singularity
 is indeed spurious,
and it can be eliminated by a mass counter term. It is surprising and
highly satisfying that the same counter terms that are needed to cancel
the ultraviolet mass divergences  also automatically cancel the infrared
singularity at $a=0$. The mean field approximation can then be
applied to the continuum limit on the world sheet, without
encountering any problems, except for a log singularity in the
coupling constant at $D=4$, which
 can be circumvented by coupling constant renormalization. The results
about soliton formation and and stringy excitations remain
unchaned, except now they are on a firmer basis.

 The continuum limit comes with an additional bonus: The model
is now invariant under the subgroup of Lorentz transformations that preserve
the light cone, including the boost $K_{1}$ along the special direction 1.
Invariance under this boost, broken when the sigma coordinate is discretized,
 is restored in the continuum limit. We will always make sure that the
approximations employed in this work  preserve this important
symmetry.

After these preliminaries, we are ready to discuss the new results
of this paper, stating with section 7. In this section, the ground
state of the model is investigated in the mean field approximation.
In this approximation,
the classical Hamiltonian, $H_{c}$, depends on
two parameters: $\lambda$ and $\rho$, or two convenient
combinations of these, $\tilde{\lambda}$ and $\tilde{\rho}$ 
(see eq.(6.1)). $\lambda$ is a Lagrange multiplier and
 $\rho$ measures the average density of the graphs on the
world sheet. The ground state energy is determined by setting
 the variation of $H_{c}$ with respect to  $\tilde{\lambda}$
 equal to zero, and then minimizing the result 
  with respect to $\tilde{\rho}$. We find two different solutions:
One of them is a trivial solution, with $\rho=0$ and an empty
world sheet. The other one has $\rho\neq 0$ and therefore a 
non-trivial world sheet populated with graphs. In this approximation,
both solutions are degenerate with vanishing
 ground state energy. Of
course, both the solitonic configurations and the resulting
string picture exist only in the non-trivial ground state.

The purpose of the present article is to establish the existence
of the non-trivial ground state and investigate some of its features.
Unfortunately, this possibility was missed in reference [1].
In retrospect, the reason for this is simple. In the light cone
set up, the $\sigma$ coordinate on the world sheet is compactified
on a circle of circumference $p^{+}$. In [1], the model was
decompactified right from the start by letting  $p^{+}\rightarrow \infty$,
and, as  explained in section 7, the non-trivial ground state is
then lost. To avoid  $\rho=0$ and the resulting empty world sheet,
one can fix $\rho$ at a non-zero value by, for example,
coupling it to fixed external source. But this is both artificial
and unnecessary; keeping the world sheet compactified avoids the loss of
the non-trivial ground state.

Having costructed the solitonic solutions, in section 8, we study the
 quantum fluctuations in the solitonic background. Here, we  focus
exclusively on a particular set of fluctuations
 that come about because the soliton, having
a definite location, breaks the translation symmetry of the model (eq.(5.5)).
It is then the standard procedure to introduce collective coordinates
corresponding to translations. Upon quantization, these collective modes
restore the spontaneously broken translation symmetry. They can therefore
be identified  as the Goldstone modes, and are expected to dominate
the low energy regime.

 In [1], it was shown that the spectrum of the
 fluctuations differ from those of
 the conventional string theory;
the Regge trajectories are no longer linear. Only in the asymptotic
limit when the density of graphs tends to infinity, the
standard string model with linear trajectories is recovered. 
In section 9, we investigate a weak coupling expansion around
 the high density limit systematically
by introducing a redefined coupling constant $\beta_{D}$ as an
 expansion parameter (eq.(9.5)). We find that in addition to the
 power series dependence on $\beta_{D}$ expected from a
perturbation expansion, there also exponential factors (eq.(9.6))
that are usually associated with tunneling. The leading term
is the usual string action in the light cone picture; the 
non-leading terms introduce corrections that tend to curve
the originally straight string trajectories. We end the section with
 a conjecture: The exponentially suppressed terms could come from the
tunneling between the two ground states. Finally, the last section summarizes
our conclusions.

\vskip 9pt
\section{The World Sheet Picture}
\vskip 9pt

The planar graphs of $\phi^{3}$ can be represented [3] on a world sheet
parameterized by the light cone coordinates $\tau=x^{+}$ and
$\sigma=p^{+}$ as a collection of horizontal solid lines (Fig.1), where
the n'th line carries a D dimensional transverse momentum $\cue_{n}$.
\begin{figure}[t]
\centerline{\epsfig{file=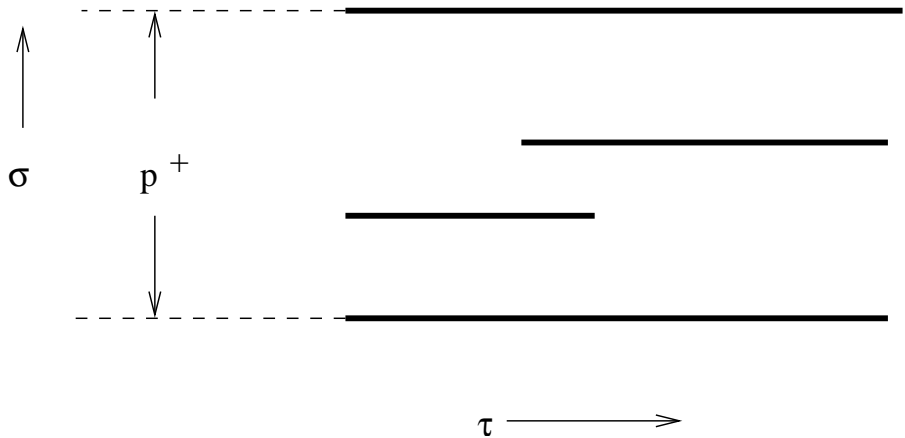, width=12cm}}
\caption{A Typical Graph}
\end{figure}
Two adjacent solid lines labeled by n and n+1 correspond to the light
cone propagator
\be
\Delta({\bf p}_{n})=\frac{\theta(\tau)}{2 p^{+}}\,\exp\left(
-i \tau\, \frac{{\bf p}_{n}^{2}+ m^{2}}{2 p^{+}}\right),
\ee
where ${\bf p}_{n}= \cue_{n}-\cue_{n+1}$ is the momentum flowing through
the propagator. A factor of the coupling constant g is inserted
 at the beginning and at the end of each line, where the interaction
takes place. Ultimately, one has to integrate over all possible
locations and lengths of the solid lines, as well as over the
momenta they carry.

The propagator (2.1) is singular at $p^{+}=0$. It is well known that 
this is a spurious singularity peculiar to the light cone picture.
To avoid this singularity  temporarily,
it is convenient to
discretize the $\sigma$ coordinate in steps of length $a$.
 A useful way of visualizing the discretized world sheet is
pictured in Fig.2. The boundaries of the propagators are marked by
solid lines as before, and the bulk is filled by dotted lines spaced
at a distance $a$.
For convenience, the $\sigma$ coordinate
 is compactified by imposing periodic
boundary conditions at $\sigma=0$ and $\sigma=p^{+}$. In contrast, the
boundary conditions at $\tau=\pm \infty$ are left arbitrary. 
 In  sections 4 and 5, it was shown how to go from a discrete
to a continuous world sheet after eliminating the
  singularity at $p^{+}=0$.

\begin{figure}[t]
\centerline{\epsfig{file=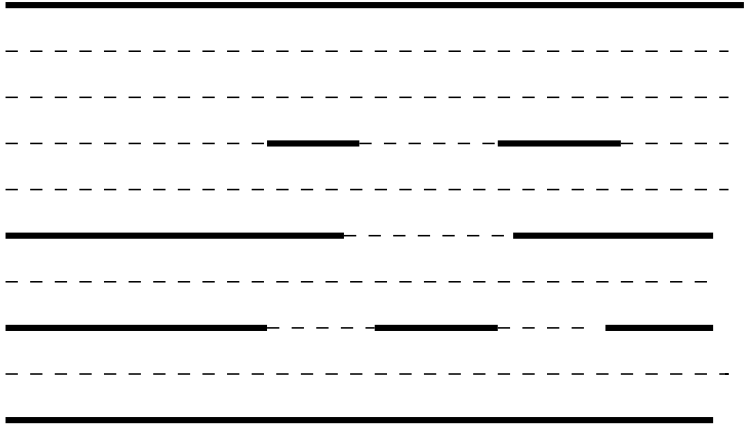, width=10cm}}
\caption{Solid And Dotted Lines}
\end{figure}

\vskip 9pt

\section{The World Sheet Field Theory}

\vskip 9pt

It was shown in [8] that the light cone graphs described above are
reproduced by a world sheet field theory, which we now briefly review.
We introduce the complex scalar field $\phi(\sigma,\tau,\cue)$ and
its conjugate $\phi^{\dagger}$, which at time $\tau$
 annihilate (create) a solid line with coordinate $\sigma$ carrying
momentum $\cue$. They satisfy the usual commutation relations
\be
[\phi(\sigma,\tau,\cue),\phi^{\dagger}(\sigma',\tau,\cue')]=
\delta_{\sigma,\sigma'}\,\delta(\cue-\cue').
\ee
The vacuum, annihilated by the $\phi$'s, represents the empty world sheet.

In addition, we introduce a two component fermion field $\psi_{i}(
\sigma,\tau)$, $i=1,2$, and its adjoint $\bar{\psi}_{i}$, which
satisfy the standard anticommutation relations. The fermion with
$i=1$ is associated with the dotted lines and $i=2$ with the solid
lines. The fermions are needed to avoid unwanted configurations
on the world sheet. For example, multiple solid lines generated by
the repeated application of $\phi^{\dagger}$ at the same $\sigma$
would lead to overcounting of the graphs. These redundant states can
be eliminated by imposing the constraint
\be
\int d\cue\, \phi^{\dagger}(\sigma,\tau,\cue)\phi(\sigma,\tau,\cue)
=\rho(\sigma,\tau),
\ee
where
\be
\rho=\bar{\psi}_{2}\psi_{2},
\ee
which is equal to one on solid lines and zero on dotted lines. This
constraint ensures that there is at most one solid line at each
site.

Fermions are also needed to avoid another set of unwanted configurations.
Propagators are assigned only to adjacent solid lines and not to
non-adjacent ones. To enforce this condition, it is convenient to
define, 
\be
\mathcal{E}(\sigma_{i},\sigma_{j})=\prod_{k=i+1}^{k=j-1}\left(
1-\rho(\sigma_{k})\right),
\ee
for $\sigma_{j}>\sigma_{i}$, and zero for $\sigma_{j}<\sigma_{i}$.
The crucial property of this function is that it acts as a projection: 
It is equal to one when the two lines at $\sigma_{i}$ and $\sigma_{j}$
are seperated only by the dotted lines; otherwise, it is zero. With the
help of $\mathcal{E}$, the free Hamiltonian can be written as
\bq
H_{0}&=&\frac{1}{2}
\sum_{\sigma,\sigma'}\int d\cue \int d\cue'\,\frac{\mathcal
{E}(\sigma,\sigma')}{\sigma'-\sigma} \left((\cue-\cue')^{2}+ m^{2}
\right)\nonumber\\
&\times& \phi^{\dagger}(\sigma,\cue) \phi(\sigma,\cue)
 \phi^{\dagger}(\sigma',\cue') \phi(\sigma',\cue')\nonumber\\
&+&\sum_{\sigma} \lambda(\sigma)\left(\int d\cue\,
 \phi^{\dagger}(\sigma,\cue) \phi(\sigma,\cue) -\rho(\sigma)\right),
\eq
where $\lambda$ is a lagrange multiplier enforcing the constraint (3.2).
The evolution operator $\exp(-i \tau H_{0})$, applied to states,
generates a collection of free propagators, without, however, the
prefactor $1/(2 p^{+})$.

One can also think of 
the Lagrange multiplier $\lambda(\sigma, \tau)$ as an Abelian
gauge field on the world sheet. The corresponding gauge
transformations are [14]
\bq
\psi &\rightarrow& \exp\left(- \frac{i}{2} \alpha\, \sigma_{3}\right)\,
\psi, \,\,\, \bar{\psi} \rightarrow \bar{\psi}\,\exp\left(\frac{i}{2}
\alpha\, \sigma_{3}\right), \nonumber\\
\phi &\rightarrow& \exp( -i \alpha)\,\phi,\,\,\,
 \phi^{\dagger} \rightarrow
\exp(i \alpha)\,\phi^{\dagger},\nonumber\\
\lambda &\rightarrow& \lambda - \partial_{\tau} \alpha.
\eq
This gauge invariance comes about because  constraint (3.2) is time
independent. Using the equations of motion,
$$
\partial_{\tau}\left(\int d\cue\,(\phi^{\dagger} \phi) -\rho\right)=0,
$$
and therefore the constraint is really needed only at a fixed $\tau$,
 say, as an initial condition. This can be implemented by
  gauge fixing by requiring $\lambda$
to be independent of the time $\tau$, 
$$
\lambda(\sigma,\tau)\rightarrow \lambda(\sigma),
$$
by a suitable choice of gauge parameter $\alpha$.  In this time
 independent form, which we will assume from now on,
$\lambda$ is not a dynamical variable but a convenient tool for 
implementing the constraint (3.2) on the initial states.

Using (3.2), the free Hamiltonian can be written in a
form more convenient for later application:
\bq
H_{0}&=&\frac{1}{2}\sum_{\sigma,\sigma'}G(\sigma,\sigma')\Bigg(
\frac{1}{2} m^{2}_{0}\,\rho(\sigma) \rho(\sigma') + \rho(\sigma')\,
\int d\cue\,(\cue^{2}+\mu^{2})
\, \phi^{\dagger}(\sigma,\cue) \phi(\sigma,\cue)\nonumber\\
&-&\int d\cue \int d\cue'\,(\cue\cdot \cue')\,
\phi^{\dagger}(\sigma,\cue) \phi(\sigma,\cue)
\phi^{\dagger}(\sigma',\cue') \phi(\sigma',\cue')\Bigg)\nonumber\\
&+&\sum_{\sigma}\lambda(\sigma)\left(\int d\cue\,
 \phi^{\dagger}(\sigma,\cue) \phi(\sigma,\cue) -\rho(\sigma)\right),
\eq
where we have defined
\be
G(\sigma,\sigma')=\frac{\mathcal{E}(\sigma,\sigma')+
\mathcal{E}(\sigma',\sigma)}{|\sigma-\sigma'|}.
\ee
There is a redundancy in the above equation: the mass is split into
two pieces according to
$$
m^{2}=m_{0}^{2}+\mu^{2}.
$$
This redundancy will prove useful later on.

Next, we introduce the interaction term. Two kinds of interaction
vertices, corresponding to $\phi^{\dagger}$ creating a solid line
or $\phi$ destroying a solid line, are pictured in Fig.3.
\begin{figure}[t]
\centerline{\epsfig{file=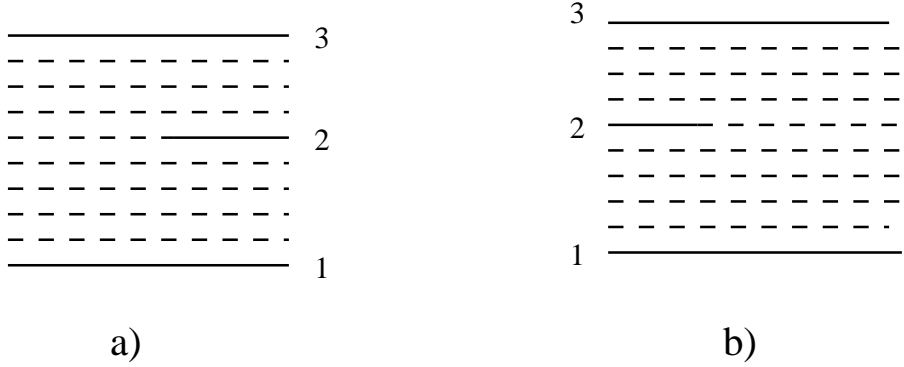, width=12cm}}
\caption{The Two $\phi^{3}$ Vertices}
\end{figure}
  
 The interaction term in the Hamiltonian, including the 
prefactors of the form $1/(p^{+})$ in (2.1), can
now be written as
\be
H_{I}= g \sqrt{a}\,\sum_{\sigma}\int d\cue\,\left(\mathcal{V}(\sigma)\,
\rho_{+}(\sigma)\, \phi(\sigma,\cue)+
\rho_{-}(\sigma)\,\mathcal{V}(\sigma)\,
\phi^{\dagger}(\sigma,\cue)\right),
\ee
where $g$ is the coupling constant, and
\be
\mathcal{V}(\sigma)=\dbsm \frac{W(\sigma_{1},\sigma_{2})}
{\sqsm},
\ee
where,
\be
\wss=\rho(\sigma_{1})\, \mathcal{E}(\sigma_{1},\sigma_{2})\,
 \rho(\sigma_{2}).
\ee
and
\be
\rho_{+}=\bar{\psi}_{1} \psi_{2},\,\,\,\rho_{-}=\bar{\psi}_{2}
\psi_{1}.
\ee

A detailed explanation of the origin of various terms in $H_{I}$
was given in [8]. 

 The total Hamiltonian is given by
\be
H=H_{0}+H_{I}
\ee
and the corresponding action by
\be
S=\int d\tau\left(\sum_{\sigma}\left(i \bar{\psi} \partial_{\tau}
\psi + i\int d\cue\,\phi^{\dagger} \partial_{\tau} \phi \right)
- H(\tau)\right).
\ee

\vskip 9pt

\section{Classical Solutions And Mass Renormalization}

\vskip 9pt

In this section,  we look for classical solutions to the
 equations motion resulting from
 the above action. However, it was
  pointed out in [9]
 that treating the $\rho$'s as classical fields is problematic. It implies
factorization of the expectation values of the products of the $\rho$'s,
which violates the spin algebras they satisfy:
\be
\rho^{2}(\sigma)=\rho(\sigma),\,\,\rho_{+}(\sigma)\rho_{-}(\sigma)= 
1-\rho(\sigma),\,\,\rho_{-}(\sigma)\rho_{+}(\sigma)=\rho(\sigma).
\ee
From the spin algebra, one can derive the overlap relations
\bq
G(\sigma,\sigma')\, \rho(\sigma')\,\rho_{-}(\sigma)\,\wss &=&\nonumber\\
\left(\delta_{\sigma',\sigma_{2}}\,\frac{1}{\sigma_{2}- \sigma}+
\delta_{\sigma',\sigma_{1}}\,\frac{1}{\sigma -\sigma_{1}}\right)
&\rho_{-}(\sigma)&\wss,
\eq
and
\be
\wss \rho_{+}(\sigma) \rho_{-}(\sigma) W(\sigma'_{1},\sigma'_{2})
=\delta_{\sigma_{1},\sigma'_{1}}\,\delta_{\sigma_{2},\sigma'_{2}}   
\,\wss.
\ee

 These overlap relations turn out to be
crucial for the elimination of both ultraviolet divergences and the
singularity at $a=0$, which is the reflection of the original
$p^{+}=0$ singularity in the propagator
 (2.1). If present, this singularity would prevent
us from taking the continuum limit of the model.

 In the classical approximation, operators are replaced by their expectation
 values. However, this violates  the  overlap relations.
To overcome this problem,  we treat
the $\phi$'s as classical fields, but keep the $\rho$'s as operators
satisfying eqs.(4.2, 4.3) in the intermediate stages of the computation.
 The  strategy is first to simplify the expressions 
 as much as possible
using the overlap relations  before making any approximations.

 We will now
search for solutions $\phi_{0}(\sigma,\cue)$ that are time independent
(solitonic) and whose dependence on $\cue$ is rotationally invariant.
The equation motion for $\phi_{0}$ then simplifies to
\be
\left(2 \lambda(\sigma)+\sum_{\sigma'} G(\sigma,\sigma')\,\rho(\sigma')
\,(\cue^{2}+\mu^{2})\right)\,\phi_{0}(\sigma,\cue) = 2 g \sqrt{a}\,
\rho_{-}(\sigma)\,\mathcal{V}(\sigma).
\ee

To solve this equation, we make the following ansatz for $\phi_{0}$:
\be
\phi_{0}(\sigma,\cue)=\dbsm\,\rho_{-}(\sigma)\,
 \wss\,\tilde{\phi}_{0}(\sigma,\sigma_{1},\sigma_{2},
\cue).
\ee
where $\tilde{\phi}$ is a c-number and all the operator dependence
is in $\rho_{-} W$. Using the overlap relations, the solution for
$\tilde{\phi}_{0}$ is given by
\bq
\tilde{\phi}_{0}(\sigma,\cue)&=&
 - \dbsm\frac{2 g\sqrt{a}}{\left(2 \lambda(\sigma)+
(\cue^{2}+\mu^{2})\left(\ssm\right)\right)}\nonumber\\
&\times&\frac{1}{\sqsm},
\eq
and the solution for $\phi_{0}^{\dagger}$ is  the Hermitian conjugate
expression.

Next, we define $H_{c}$ by replacing $\phi$ by the above $\phi_{0}$
 in the Hamiltonian,
$$
H_{c}=H(\phi=\phi_{0}),
$$
 and simplify again using the overlap relations until
we have a linear result in $W$:
\bq
H_{c}&=& -2 g^{2} a\,\sum_{\sigma} \dbsm\,\int d\cue\, \wss
\nonumber\\
&\times& \left(\qsm\,\left(2 \lambda(\sigma)+
 (\cue^{2} +\mu^{2}) \ssm\right)\right)^{-1}\nonumber\\
&-&\sum_{\sigma} \lambda(\sigma)\,\rho(\sigma)+ \frac{m_{0}^{2}}{2}
\sum_{\sigma'>\sigma}\frac{W(\sigma,\sigma')}{\sigma' -\sigma}.
\eq

In the above expression, the integral over $\cue$ is ultraviolet divergent
at $D=2$ and $D=4$.
This divergence can be eliminated by the mass renormalization
 and at $D=4$ by also  coupling
constant renormalization. We observe that as $|\cue|\rightarrow \infty$,
the first term on the right, after doing the sum over $\sigma$, reaches
a limit identical in form to the mass term. It can therefore be cancelled by
setting
\be
m_{0}^{2}= 4 g^{2} a\,\int d\cue\,\frac{1}{\cue^{2}+\mu^{2}}.
\ee
We note that at $D=2$, there is no
divergence, and at $D=4$, a quadratic divergence is reduced to a logarithmic
divergence in the coupling constant. Although there is no divergence at
$D=1$, we will still use the same expression for $m_{0}$ also in this case.

At the beginning, we started with two independent masses in the
problem. But now  that $m_{0}$ is fixed, only
$\mu$  remains. We could have given a treatment based on a single mass
from the start, however,
having an extra mass temporarily is more convenient. For example, it
 enables us to give a uniform
treatment for all dimensions.

Up to this point, the world sheet is still
discrete, and the continuum limit $a\rightarrow 0$ is problematic. 
This problem will be
addressed in the next section.

\vskip 9pt

\section{The Continuum Limit}

\vskip 9pt

The continuum limit is taken by letting $a\rightarrow 0$, after suitably
scaling the field variables by
\be
\phi\rightarrow \sqrt{a}\,\phi,\,\,\,\psi\rightarrow \sqrt{a}\,\psi.
\ee
 From its definition, $\rho$ scales as
\be
\rho\rightarrow a\,\rho.
\ee
 In this limit,
all the sigma sums become integrals, and all the factors of $a$ are used up
in this process.
Also, the product in the
definition of $\mathcal{E}$ (3.4) becomes 
\be
\mathcal{E}(\sigma_{1},\sigma_{2})=\prod_{\sigma_{1}}^{\sigma_{2}}
\left(1- a\,\rho(\sigma)\right)
\rightarrow \exp\left(- \int_{\sigma_{1}}
^{\sigma_{2}} d\sigma\,\rho(\sigma)\right).
\ee

 After a change of variables by
$$
\sigma=\sigma_{1}+x\,(\sigma_{2} -\sigma_{1}),
$$
$H_{c}$ can be written as
\bq
H_{c}&=& -2 g^{2} \int d\sigma_{2} \int^{\sigma_{2}} d\sigma_{1}
\int_{0}^{1} dx\int d\cue\, \rho(\sigma_{1})\,
\mathcal{E}(\sigma_{1},\sigma_{2})\,\rho(\sigma_{2})\nonumber\\
&\times &
\left((\sigma_{2} -\sigma_{1})
\left(2 \lambda(\sigma)\,x (1-x)(\sigma_{2} -\sigma_{1})+ 
 (\cue^{2} +\mu^{2})\right)\right)^{-1}\nonumber\\
&+& 2 g^{2}\,\int d\sigma_{2}
\,\int^{\sigma_{2}} d\sigma_{1}
\int d\cue\,\frac{1}{\cue^{2}+\mu^{2}}
\,\frac{\rho(\sigma_{1})\,
\mathcal{E}(\sigma_{1},\sigma_{2})\,\rho(\sigma_{2})}{\sigma_{2} -\sigma_{1}}
- \int d\sigma\, \lambda(\sigma)\,\rho(\sigma).\nonumber\\
&&
\eq

The first and the second terms on the right
 are divergent as $|\cue|\rightarrow \infty$ at
$D=2, 4$, and also they are also logarithmically divergent as
$\sigma_{2}-\sigma_{1}\rightarrow 0$. The first is the ultraviolet mass
 divergence and we have already fixed $m_{0}$ by eq.(4.8) so that
it cancels between the two terms.
 The second singularity is a logarithmic singularity at
$\sigma_{2}-\sigma_{1}=0$. Since $\sigma_{2}-\sigma_{1}$ is the $p^{+}$
 flowing through the propagator, this is 
 the $p^{+}=0$ singularity in disguise. 
Surprisingly, this divergence also cancels between the first and second terms
 in all dimensions. It is highly satisfying that the mass counter term
introduced to eliminate an ultraviolet divergence also automatically
cancels the infrared divergence at $p^{+}=0$. This cancellation is quite
non-trivial and absolutely essential, since otherwise, having 
only one adjustable constant $m_{0}$ at our disposal, 
 we would be stuck with one divergence or other at  $D=2, 4$. We also
note that we cannot add an arbitrary ultraviolet
 finite term to $m_{0}^{2}$ without spoiling the infrared cancellation.
Although we started with two masses, in the end only $\mu$ remains as
an arbitrary parameter.

Another important feature of $H_{c}$ is its symmetries. In addition to
translation invariance in $\cue$ 
\be
\cue\rightarrow \cue+{\bf r},
\ee
the light cone dynamics is manifestly invariant under a subgroup of Lorentz
transformations. The original action (3.14) is trivially invariant under
under all the generators of this subgroup except for  the generator
$K_{1}$ of boosts along the special direction $1$. The discretization
of the $\sigma$ coordinate breaks this symmetry even at the classical level.
We expect this symmetry will be at least classically
 restored in the continuum limit. To see
this, we note that under $K_{1}$, various fields transform as
\bq
\phi(\sigma,\tau,\cue)&\rightarrow& \sqrt{u}\,\phi(u \sigma, u \tau,\cue),
\,\,\psi(\sigma,\tau,)\rightarrow \sqrt{u}\,\psi(u \sigma, u \tau),\nonumber\\
\rho(\sigma,\tau)&\rightarrow& u\,\rho(u \sigma, u \tau),\,\,
\lambda(\sigma, \tau)\rightarrow u\,\lambda(u \sigma, u \tau),\,\,
p^{+}\rightarrow \frac{1}{u}\,p^{+},
\eq
where $u$ parametrizes the $K_{1}$ transformations. In the expression for
$H_{c}$, this amounts to letting
$$
\sigma\rightarrow u\,\sigma,\,\,\,\tau\rightarrow u\,\tau,
$$
and transforming $\rho$ according to eq.(31). The classical Hamiltonian then
transforms as
\be
H_{c}\rightarrow u\,H_{c},
\ee
and as expected,
the corresponding action is therefore invariant. As we shall see, this
 invariance will be respected by the mean field approximation, and it
will play an important role in what follows.

Eq.(5.4), which is free of divergences and  independent of $a$, will be
the starting point of the mean field approximation in the next section.

\vskip 9pt

\section{The Meanfield Approximation}

\vskip 9pt

The mean field approximation consists of replacing $\rho$ and $\lambda$
in $H_{c}$ by their ground state expectation values, which we assume to be
independent of $\sigma$ and $\tau$.
(translation invariance of the ground state).
 Afterwards,    the equation
of motion with respect to the
gauge fixed  $\lambda$ should be imposed
 as a constraint, and  the resulting
$H_{c}$ should be minimized with respect to $\rho$ to
find the ground state. We remind the reader that this is the standard
procedure in fixing an axial gauge:  The equations
of motion with respect to gauge fixed variable are imposed as
constraints.

In  eq.(5.4), the $\cue$ integration
 can be done, and the result can be simplified by
the following change of variables:
\be
\tilde{\lambda}= \lambda/(\rho \mu^{2}),\,\,\,\sigma= \sigma'/\rho,
\,\,\,\tilde{\rho}=\rho p^{+}.
\ee
These variables have advantage of being both invariant under
$K_{1}$ and scale independent. Also $\tilde{\rho}$ is a physically
significant variable; it counts the number of solid lines and hence
the number of propagators on the world sheet. $p^{+}$ and $\rho$
seperately are not physically meaningful: They depend on the choice
of the Lorentz frame since they are not $K_{1}$ invariant.

 In terms of these new variables,
the classical Hamiltonian for various transverse dimensions $D$
 can then be written as
\be
p^{+}\,H_{c}= \tilde{\rho}^{2}\,F_{D}(\tilde{\lambda},\,\tilde{\rho}),
\ee
where,
\be
F_{D}=\mu^{2}\,\left(
 -\tilde{\lambda} +\alpha_{D}\,\int_{0}^{\tilde{\rho}} d\sigma'
\int_{0}^{1} d x\, \frac{\exp(-\sigma')}{\sigma'}\
 L_{D}(x,\sigma',\tilde{\lambda})\right),
\ee
with
\be
\alpha_{1}= 2\pi g^{2}/\mu^{3},
\,\,\alpha_{2}= 2\pi g^{2}/\mu^{2},
\,\,\alpha_{4}=2 \pi^{2} g^{2},
\ee
and,
\bq
 L_{1}&=& 1 -
\frac{1}{\left(1+ 2\tilde{\lambda}\,x (1-x)
\sigma'\right)^{1/2}},\nonumber\\
 L_{2}&=&\ln\left(1+ 2\tilde{\lambda}\,x (1-x)\,\sigma'
\right),\nonumber\\
L_{4}&=&2\tilde{\lambda}\,x (1-x)\,\sigma'\,\ln\left(\Lambda^{2}/
\mu^{2}\right)\nonumber\\
&-&\left(1+2\tilde{\lambda}\,x (1-x)\,\sigma'
\right)\,\ln\left(1+2\tilde{\lambda}\,x (1-x)\,\sigma'
\right).
\eq

In the last equation, $\Lambda$ is an ultraviolet cutoff. These equations fix
$H_{c}$ in terms of dimensionless coupling constants
$\alpha_{1,2}$ at $D=1,2$. At $D=4$,
the expression for $L_{4}$ has a logarithmic dependence on the
cutoff $\Lambda$. This is related to  coupling constant renormalization.
 We recall that $\phi^{3}$ is asymptotically free
in 6 space-time dimensions ($D=4$),
 and the above relation is the well known lowest order renormalization
group result obtained by summing the leading logarithmic
divergences in the perturbation series. To get a finite result,
 one should first
renormalize the coupling constant before summing the logs. This
amounts to replacing the cutoff $\Lambda$ by a large but finite
value. The coupling constant on the left should then be identified
with  the running coupling constant $g(\Lambda)$, defined at the energy
scale  $\Lambda$. For this leading log. approximation to be reliable,
$g(\Lambda)$ should be small, which means that $\Lambda^{2}/\mu^{2}$
should be large. All the additional terms on the right hand side only
make a small change in the scale
of the running coupling constant. From now on, we will only keep
the leading first term for  $L_{4}$.

\vskip 9pt

\section{The Ground State}

\vskip 9pt

We will now investigate the ground state of the model in various dimensions,
using the meanfield approximation developed in the last section. We remind
the reader that $\tilde{\lambda}$ and $\tilde{\rho}$ are taken to be constants
independent of $\sigma$ and $\tau$, and the equation
\be
\frac{\partial H_{c}}{\partial \tilde{\lambda}}=0
\ee
is imposed as a constraint. Since $H_{c}$ is proportional to 
$\tilde{\rho}^{2}$,
 this equation always has the trivial solution
\be
\tilde{\rho}=0,\,\, H_{c}=0.
\ee
This corresponds to an uninteresting empty world sheet.

We will now show that there is another more interesting solution with
$$
\tilde{\rho}\neq 0.
$$
and with again
$$
H_{c}=0.
$$
This non-trivial ground state, degenerate in energy with the trivial one,
correponds to a
 world sheet populated with Feynman graphs. This solution
is obtained by setting
\be
\frac{\partial F_{D}}{\partial \tilde{\lambda}}= 0.
\ee

We will now study this equation for various $D$.
Starting with $D=1$, it reduces to 
\be
\alpha_{1}\,\int_{0}^{\tilde{\rho}} d\sigma' \int_{0}^{1} dx
\,\frac{x (1-x)\,\exp(- \sigma')}{\left(1+ 2 \tilde{\lambda}\,
x (1-x)\,\sigma'\right)^{3/2}}= 1.
\ee

Now a few comments:\\
a) Because of this constraint, we are left with only one
independent variable, which we take to be $\tilde{\rho}$.
$\tilde{\lambda}$ is treated as a function of $\tilde{\rho}$.
\\
b) Both $\tilde{\rho}$ and $\tilde{\lambda}$ are positive semi-definite;
 the first by definition and the other by virtue of
the above equation.\\
c) The left hand side is an increasing function of $\tilde{\rho}$
and a decreasing function of $\tilde{\lambda}$. It is then easy to see
that the minimum value of  $\tilde{\lambda}$,
\be
\tilde{\lambda}=0,
\ee
corresponds also to the minimum value of $\tilde{\rho}$, which 
we label $\rho^{c}_{1}$ (1 refers to D). Solving (7.4) for
$\tilde{\rho}$ at $\tilde{\lambda}=0$, we have,
\be
 \rho^{c}_{1}=
-\ln\left(1-\frac{6}{\alpha_{1}}\right).
\ee
 For this solution to exist, $\alpha_{1}$ must satisfy
\be
\alpha_{1}>6. 
\ee
Clearly, this corresponds to the strong coupling regime.\\
d) $\tilde{\lambda}$ is a monotonically increasing function of
$\tilde{\rho}$. As $\tilde{\rho}$ ranges from $\rho^{c}_{1}$ to
$\infty$,  $\tilde{\lambda}$ ranges from 0 to $\infty$.

Next, we show that $F_{1}$ is also
 a monotonically increasing function $\tilde{\rho}$, and therefore,
 its minimum is at $\tilde{\rho}= \rho^{c}_{1}$, the minimum
value of  $\tilde{\rho}$. Differentiating
$F_{1}$ (eq.(6.3)) with respect to $\tilde{\rho}$ and remembering
that $\tilde{\lambda}$ is a function of $\tilde{\rho}$ through
eq.(7.4), we have
\be
\frac{d F_{1}}{d \tilde{\rho}}=\mu^{2}\,\alpha_{1}\,
\frac{\exp(-\tilde{\rho})}{\tilde{\rho}}\,
\int_{0}^{1} d x\,
\left(1 - \left(1+2 \tilde{\lambda}\,x (1-x)\,\tilde{\rho}
\right)^{- 1/2}\right).
\ee
Since the right hand side is positive for
$$
\tilde{\lambda}>0,
$$
it follows that
$$
\frac{d F_{1}}{d \tilde{\rho}}>0
$$
for
$$
\tilde{\rho}>\rho^{c}_{1}.
$$
 Finally, it is easy to show that since
$\tilde{\lambda}$ vanishes at $\tilde{\rho}=\rho^{c}_{1}$,
both $F_{1}$ and its derivative with respect to $\tilde{\rho}$
vanish at the same point.

 Having shown that $F_{1}(\tilde{\rho})$ has a global
minimum at
$$
\tilde{\rho}=\rho^{c}_{1},
$$
with
\be
F_{1}(\rho^{c}_{1})=0,
\ee
we will now show that
$H_{c}$ also has a vanishing minimum at the same point.
From eq.(6.2),
 $p^{+}\,H_{c}$ is the product of $F_{1}(\tilde{\rho})$            
 and $\tilde{\rho}^{2}$. Since both factors reach their minimum at
$\tilde{\rho}=\rho^{c}_{1}$,  $H_{c}$ also reaches its minimum value
zero at the same point. Being a global minimum, this corresponds to
a stable ground state within the parameter space we have been considering.
Of course, this is only a classical result; quantum fluctuations
could destabilize it.

Next, we consider $D=2$, which can be treated in exactly same fashion
as $D=1$, with only some obvious minor changes. Eq.(7.4) is now replaced
by
\be
\alpha_{2}\,\int_{0}^{\tilde{\rho}} d\sigma' \int_{0}^{1} dx
\,\frac{x (1-x)\,\exp(- \sigma')}{1+ 2 \tilde{\lambda}\,
x (1-x)\,\sigma'}= 1.
\ee
We can repeat the argument following eq.(7.4), with the only change that
the minimum value of $\tilde{\rho}$ is now
\be
\rho^{c}_{2}=-\ln\left(1-\frac{6}{\alpha_{2}}\right),
\ee
and for a solution to exist, $\alpha_{2}$ must be greater than 6.

The results following (7.4) are still valid, but eq.(7.8) is now replaced by
\be
\frac{d F_{2}}{d \tilde{\rho}}=\mu^{2}\,\alpha_{2}\,
\frac{\exp(-\tilde{\rho})}{\tilde{\rho}}\,
\int_{0}^{1} d x\,
\ln \left(1+2 \tilde{\lambda}\,x (1-x)\,\tilde{\rho}
\right).
\ee
From this equation, one can easily show that, replacing
$F_{1}$ by $F_{2}$, the argument following (7.9)
is still valid, and therefore
$$
\tilde{\rho}=\rho^{c}_{2}
$$
corresponds to a stable classical ground state.

Finally, we will briefly discuss the $D=4$ case. Because of the running
 coupling constant, there are additional complications compared to
$D=1, 2$, and our treatment will be less complete. Eq.(7.3) at D=4
gives
\be
\tilde{\rho}=
\rho^{c}_{4}= -\ln\left(1-\frac{3}{\bar{\alpha}_{4}}\right),
\ee
where
$$
\bar{\alpha}_{4}=\alpha_{4}\,\ln\left(\Lambda^{2}/
\mu^{2}\right),
$$
and we have kept only the leading log term. In this case, since
$H_{c}$ is linear in $\lambda$, this variable acts as a Lagrange
multiplier, $\tilde{\rho}$ is fixed at $\rho^{c}_{4}$, and no
fluctuations are allowed.
$\lambda$ remains arbitrary, and the classical energy is again zero.
Although we will not pursue it further here, higher order
corrections could easily change this picture.

We now return to the question of why  the
non-trivial ground state corresponding to $\tilde{\rho}\neq 0$ was
missed in reference [1]. As explained in the Introduction, 
this was because, in [1],
only the decompactified model, with
$$
p^{+}\rightarrow \infty,
$$
and consequently,
$$
\tilde{\rho}\rightarrow \infty
$$
was studied.
Actually, $\tilde{\rho}\rightarrow \infty$ is not a solution for the
ground state, but the
asymptotic limit of the ground states  described by eqs.(7.6), (7.11)
and (7.13) as $\alpha_{D}$ tends to its limiting values
\be
\alpha_{1, 2}\rightarrow 6
\ee
for $D=1, 2$ and,
\be
\bar{\alpha}_{4}\rightarrow 3,
\ee
for $D=4$. By setting $\tilde{\rho}=\infty$ from the very beginning,
this subtle point was missed in [1]. We will study this interesting
asymptotic limit, which we call the high density limit,
 in the following sections.

\vskip 9pt

\section{Fluctuations Of The Transverse Momentum 
 Around The Classical Background} 

\vskip 9pt

Given the classical solutions developed in the previous sections, it is
natural  to study quantum fluctuations about these
backgrounds. This can be done explicitly to  quadratic order for all the
 fluctuations. We will, instead, focus  on a particular
set of fluctuations; namely, the fluctuations of the transverse momentum
$\cue$, which can be studied
 by quantizing the collective coordinates corresponding
to the breaking of the translation invariance of $\cue$ (eq.(5.5)).
The classical solution, placed at a definite
location in the $\cue$ space, breaks this symmetry, and 
it is restored by quantizing the so-called collective modes.
These modes
are very important not only
for their role in restoring translation
invariance, but also, because, they are the low lying Goldstone modes
connected with the spontaneously broken translation symmetry.
Also, they were crucial to the
formation  of a string on the world sheet.

The collective coordinate corresponding to translations is
introduced by letting
\be
\phi=\phi_{0}+\phi_{1},
\ee
where $\phi_{1}$ is the fluctuating part of the field, and setting,
\be
\phi_{1}(\sigma,\tau,\cue)=
 \phi_{0}(\sigma,\cue+\bfv(\sigma,\tau))
-\phi_{0}(\sigma,\cue),
\ee
where $\phi_{0}$ is the classical solution and
$\bfv$ is the collective coordinate. The contribution
of $\phi_{1}$ to the action
can be written as the sum of kinetic and potential terms:
\be
S^{(1)}= S_{k.e}- \int d\tau\,H_{0}(\phi_{1})=S_{k.e}+ S_{p.e},
\ee
where the kinetic term depends on $\partial_{\tau}\bfv$  and the potential
has no $\tau$ derivatives. We note that only $H_{0}$ contributes
 to $S_{p.e}$; so
 substituting the ansatz (8.2) directly into $H_{0}$ (eq.(3.7)) and
 simplifying, we have the following result for all D:
\be
 S_{p.e}= -\frac{1}{4}\int d\tau \int d\sigma \int d\sigma'\,
\frac{W(\sigma,\sigma')}{|\sigma-\sigma'|}\left(\bfv(\sigma,\tau)
- \bfv(\sigma',\tau)\right)^{2}.
\ee

We note that so far no approximation was made, and therefore, this
result is exact so long as only the contribution of the collective
coordinate $\bfv$ is concerned. Also,
 there is no singularity at $\sigma=\sigma'$ and so there
is no obstacle to taking the continuum limit immediately. At this point,
 we introduce the
mean field approximation by setting  $\rho p^{+}=\rho^{c}_{D}$, its
ground state value, and change variables by
$$
\sigma'=\sigma+ z\,p^{+}.
$$
\be
 S_{p.e}\rightarrow -\frac{\left(\rho^{c}_{D}\right)^{2}}{2 (p^{+})^{2}}
 \int d\tau \int_{0}^{p^{+}} d\sigma 
\int_{0}^{1} dz\, \frac{\exp\left(-\rho^{c}_{D}\,z
\right)}{z}
\left(\bfv(\sigma+z\,p^{+},\tau)- \bfv(\sigma,\tau)\right)^{2}.
\ee

We will study this action in detail later on, but before that, we turn
our attention to the kinetic energy term. To compute this
term to quadratic order in $\partial_{\tau}\bfv$,
 one has to split $\phi_{1}$
into its real and imaginary (Hermitian and anti-Hermitian) parts:
\be
\phi_{1}= \phi_{1,r}+\phi_{1,i},
\ee
and eliminate one of them by integrating over it. In this case, since
the classical solution $\phi_{0}$ is real, 
 $\phi_{1,i}$ will be integrated out. The kinetic energy term in the
action (3.14) can then be rewritten as
\bq
&& i \sum_{\sigma} \int d\tau
\int d\cue\,\phi^{\dagger} \partial_{\tau} \phi= 2 \sum_{\sigma}
 \int d\tau \int d\cue\,\phi_{1,i}\,\partial_{\tau}\phi_{1,r}
\nonumber\\
&\rightarrow& 2 \sum_{\sigma}\int d\tau 
\int d\cue\,\phi_{1,i}\,\partial_{\tau}
 \phi_{0}(\sigma,\cue+\bfv(\sigma,\tau)).
\eq

 Integrating 
over  $\phi_{1,i}$ then amounts to solving the equations of motion for
 $\phi_{1,i}$ and substituting in the action. The left hand side of the
equation of motion is the same as in (4.4), but the right hand side comes from
the  variation of the above kinetic term  with respect to  $\phi_{1,i}$:
\be
\left(2 \lambda(\sigma)+\sum_{\sigma'} G(\sigma,\sigma')\,\rho(\sigma')
\,(\cue^{2}+\mu^{2})\right)\,\phi_{1,i}(\sigma,\tau,\cue)=
2\partial_{\tau}\phi_{0}(\sigma,\cue+\bfv(\sigma,\tau)).
\ee
This equation can be solved by letting
\be
\phi_{1,i}(\sigma,\tau,\cue)=\dbsm\,\rho_{-}(\sigma)\,
 \wss\,\tilde{\phi}_{1.i}(\sigma,\sigma_{1},\sigma_{2},\tau,
\cue),
\ee
as in (4.5).
Following the same steps  as before, this can then be simplified using
the overlap relations, and after some algebra, we have the solution
\be
\tilde{\phi}_{1.i}(\sigma,\sigma_{1},\sigma_{2},\tau)=
\frac{2\, \partial_{\tau}\bfv(\sigma.\tau)\cdot{\bf \bigtriangledown}_{q}
\tilde{\phi}_{0}(\sigma,\sigma_{1},\sigma_{2},\tau,\cue)}
{\left(2\, \lambda(\sigma)+
(\cue^{2}+\mu^{2})\left(\ssm\right)\right)},
\ee
where $\tilde{\phi}_{0}$ is given by (4.6). It is now easy to take
 the continuum limit, and
apply the mean field approximation by replacing $\lambda$ and
$\rho$ by their ground state values
$$
\lambda\rightarrow 0,\,\,\,\rho\rightarrow \rho^{c}_{D}/p^{+}.
$$
 We skip the intermediate steps give the final result for only
$D=1,2$:
\be
S_{k.e}=\int d\tau \int_{0}^{p^{+}} d\sigma\,
 \frac{1}{2}\,E(\rho^{c}_{D})\,
\left(\partial_{\tau} \bfv(\sigma,\tau)\right)^{2},
\ee
where,
\bq
E&=&\frac{128}{D}\,g^{2}\,\int_{0}^{p^{+}} dy \,\int_{0}^{y}
dx\int d{\cue}\,\frac{x^{2}\,(y-x)^{2}\,(\rho^{c}_{D})^{2}\,{\cue}^{2}}
{y^{4}\,({\cue}^{2}+\mu^{2})^{5}}\,\exp\left(-\rho^{c}_{D}
\,y\right)
\nonumber\\
&=&\frac{\alpha_{D}\,C_{D}}
{\mu^{4}}\,\left(1-(1+\rho^{c}_{D})
\exp(-\rho^{c}_{D})\right),
\eq
and,
$$
C_{1}=\frac{1}{12},\,\,\,C_{2}=\frac{4}{45}.
$$

\vskip 9pt

\section{String Formation In The High Density Limit}

\vskip 9pt

In this section, we will study the spectrum of the collective coordinate
$\bfv$, with the action  given by the
sum of $S_{p.e}$ (eq.(8.4)) and  $S_{k.e}$ (eqs.(8.11, 8.12)).
 This is a free
field theory and therefore it is exactly solvable. In fact, without any
further approximations, the spectrum
of  $S_{p.e}$ was determined in [1].
Here, we will only
 consider the  high density
 (large $\rho^{c}_{D}$) limit, which corresponds to
the coupling constants approaching the bound given by eq.(7.15).
In this limit, we expand the term involving $\bfv$
in eq.(8.5) in powers of $z$ (derivative expansion):
\be
\bfv(\sigma+z\,p^{+},\tau)-\bfv(\sigma,\tau)= z\,p^{+}\,
\partial_{\sigma} \bfv(\sigma,\tau)+\cdots.
\ee
Keeping only the leading term in the expansion and adding
the kinetic energy term  results in the action
\bq
S^{(1)}&\rightarrow& \frac{1}{2}\,\left(1-(1+\rho^{c}_{D})
\exp(-\rho^{c}_{D})\right)\nonumber\\
&\times&\int d\tau \int_{0}^{p^{+}}
\left(\frac{\alpha_{D}\,C_{D}}{\mu^{4}}\,
\left(\partial_{\tau} \bfv(\sigma,\tau)
\right)^{2}
-\left(\partial_{\sigma} \bfv(\sigma,\tau)
\right)^{2}\right).
\eq

This is the string action in the lightcone picture with the slope
$$
\alpha'=\left(\frac{\alpha_{D}\,C_{D}}{2 \pi^{2}\,\mu^{4}}\right)^{1/2}.
$$
We would like to emphasize that only the leading term was kept
 in the derivative expansion; the inclusion of the
 higher order terms produces
deviations from the string picture by introducing higher 
derivatives in $\sigma$ which tend to curve the string trajectories. 
 To see this, we exhibit the next order
term:
\bq
S_{p.e}^{(2)}&=&-\frac{(p^{+})^{2}}{8 (\rho^{c}_{D})^{2}}\,
\left(6 -\left(6+ 6 \rho^{c}_{D}+3 (\rho^{c}_{D})^{2}+
(\rho^{c}_{D})^{3}\right)\,\exp(-\rho^{c}_{D})\right)\nonumber\\
&\times&\int d\tau \int_{0}^{p^{+}} d\sigma\,\left(\partial_{\sigma}^{2}
\bfv(\sigma, \tau)\right)^{2}.
\eq

Let us compare this to the leading term (eq.(9.2)), neglecting terms
exponentially suppressed in $\rho^{c}_{D}$. We have
\be
S_{p.e}^{(2)}\approx \frac{3 (p^{+})^{2}}{2 \,(\rho^{c}_{D})^{2}}\,
S_{p.e}^{(1)}.
\ee
We note the two additional derivatives in $S_{p.e}^{(2)}$
compared to $S_{p.e}^{(1)}$ and the extra factor of
$$
\frac{(p^{+})^{2}}{(\rho^{c}_{D})^{2}}
$$
on the right. In fact, it is easy to show that, apart from numerical
 factors,  each extra
derivative with respect to $\sigma$ goes with a factor of
$p^{+}/\rho^{c}_{D}$. The factor of $p^{+}$ is needed for invariance
under $K_{1}$, and $1/\rho^{c}_{D}$ can be associated with a
perturbative expansion in a new coupling constant
$\beta_{D}$ defined by
\be
\rho^{c}_{D}=1/\left(\beta_{D}\right)^{2}.
\ee
 An expansion in powers of
  $\left(\beta_{D}\right)^{2}$  coincides with the
derivative expansion around the high density limit $\rho^{c}_{D}=0$.

Such a perturbative treatment of the model is an attractive
possibility; however, we have 
now consider the so far neglected exponential factor
\be
\exp(-\rho^{c}_{D})=\exp\left(-\frac{1}{(\beta_{D})^{2}}\right).
\ee
This clearly not perturbative but looks very much like the
tunneling factors familiar from  instanton calculations.
What is missing is the physical picture of tunneling:
 Between which two or more states
 does the tunneling take place? An natural conjecture is to
identify these states with the two (trivial and non-trivial)
ground states. However, so far we have not been able to
construct an instanton configuration that connects them.

\vskip 9pt

\section{Conclusions}

\vskip 9pt

As  emphasized in the introduction, the present paper supplements
reference [1] by providing an important missing step. The main 
contribution of that reference was a singularity free treatment of scalar
$\phi^{3}$ on the world sheet. In particular, by eliminating
the singularity at $p^{+}=0$, the original discretized world sheet
could be replaced by a continuous one. What was missing was a complete
treatment of the ground state of the model. As explained in the text,
this was due to an unnecessary decompactification of the $\sigma$
coordinate on the world sheet. By keeping the model compactified,
we show here that there are
 two ground states: One of them corresponds to
a trivial empty world sheet, and the other to a non-trivial
populated world sheet. They are degenerate at zero energy.

In [9, 10, 1], it was shown that a densely populated world sheet leads to
 string formation. To investigate this high density limit more
syatematically, we consider an expansion in terms of a redefined
coupling constant, and show that the leading term in this expansion
reproduces the light cone string action. This expansion has the
intersting feature that, in addition to the usual perturbative
terms, it has also exponentially suppressed terms. We speculate
that these terms may arise from tunneling between the two
ground states.

\vskip 9pt

\noindent{\bf Acknowledgement}

\vskip 9pt

This work was supported in part by the director, Office of Science,
Office of High Energy Physics of the U.S. Department of Energy under Contract
DE-AC02--05CH11231.

\vskip 9pt

\noindent{\bf References}

\vskip 9pt

\begin{enumerate}

\item K.Bardakci, JHEP {\bf 1306} (2013) 066, arXiv:1304.1466.
\item K.Bardakci and C.B.Thorn, Nucl.Phys. {\bf B 626} (2002)
287, hep-th/0110301.
\item G.'t Hooft, Nucl.Phys. {\bf B 72} (1974) 461.
\item C.B.Thorn, Nucl.Phys. {\bf B 637} (2002) 272, hep-th/0203167,
S.Gudmundsson, C.B.Thorn and T.A.Tran, Nucl.Phys. {\bf B 649} 92003)
3-38, hep-th/0209102.
\item C.B.Thorn and T.A.Tran, Nucl.Phys. {\bf B 677} (2004) 289,
hep-th/0307203.
\item H.P.Nielsen and P.Olesen, Phys.Lett. {\bf B 32} (1970) 203.
\item B.Sakita and M.A.Virasoro, Phys.Rev.Lett. {\bf 24} (1970) 1146.
\item K.Bardakci, JHEP {\bf 0810} (2008) 056, arXiv:0808.2959.
\item K.Bardakci, JHEP {\bf 1110} (2011) 071, arXiv:1107.5324.
\item K.Bardakci, JHEP {\bf 0903} (2009) 088, arXiv:0901.0949.

\end{enumerate}

\end{document}